\documentclass[a4paper,11pt]{article}
\usepackage{jcappub}
\usepackage{graphicx}
\usepackage[dvipsnames]{xcolor}
\usepackage{amsmath}
\usepackage{siunitx}
\DeclareSIUnit\PeV{PeV} 
\DeclareSIUnit\EeV{EeV}  
\DeclareSIUnit\Mpc{Mpc}  
\DeclareSIUnit\yr{yr}  
\DeclareSIUnit\yrs{yrs} 
\DeclareSIUnit\erg{erg} 

\title{Cosmogenic photon fluxes at ultra-high energies}
\author{Marcus Niechciol, Chiara Jane Papior, Markus Risse}

\affiliation{Center for Particle Physics Siegen, Department of Physics, University of Siegen, Walter-Flex-Str. 3, 57072 Siegen, Germany}

\emailAdd{niechciol@physik.uni-siegen.de}
\emailAdd{papior@hep.physik.uni-siegen.de}
\emailAdd{risse@hep.physik.uni-siegen.de}

\abstract{During their propagation, ultra-high energy cosmic rays produce cosmogenic photons.
The expected flux level of these photons may vary by orders of magnitude depending on parameters such as the spectrum and composition of cosmic rays at injection or the source distance.
We investigate the photon yields for various assumptions on injection parameters.
The photon yield is largest for proton-emitting sources at about $\SI{15}{Mpc}$ distance for cosmogenic photons above $\SI{e18}{eV}$. While the photon yield from protons always exceeds the one from heavier nuclei of the same total energy, the differences are reduced for larger source distances.
Then, we quantify the cosmogenic photon fluxes for different source scenarios.
We regard mixed-composition scenarios that were found to provide a reasonable description of the data from the Pierre Auger Observatory.
In addition, benchmark scenarios leading to comparably high (``maximum'') or low (``minimum'') photon fluxes are determined assuming pure proton or pure iron primaries.
For the mixed-composition scenarios that do not contain initial protons, the predicted cosmogenic photon fluxes are below present experimental limits by more than 1.5 orders of magnitude.
In case of a substantial fraction of protons at the highest energies, the related photon flux might be in reach.
Certain parameter combinations of pure proton scenarios are constrained already by present photon limits.
}

\keywords{Cosmogenic Photons, GZK Photons, Ultra-High Energy Cosmic Rays, Propagation}

\begin{document}

\maketitle

\section{Introduction}

Giant air-shower observatories are sensitive to photons at ultra-high energy (UHE, here $>\SI{e17}{eV}$)~\cite{Risse:2007sd}.
Such photons can be produced in different ways.
In exotic scenarios, by decay or annihilation of speculative new matter, a substantial fraction of energy can be transferred to photons.
However, the limits derived by air-shower observatories place severe constrains on such scenarios~\cite{PierreAuger:2022gkb}.
In conventional scenarios, where nuclei are accelerated, photons could be generated near the sources by collisions with matter or intense radiation fields (``beam dump'').
Due to our limited knowledge concerning the sources and their environments, it is unclear to what extent such photons are created and can escape the source region.

Another production channel comes from the propagation of nuclear cosmic rays, generating cosmogenic photons.
The main interaction is by photo-pion production of cosmic rays in collisions with extragalactic background fields,
in particular with the CMB, for nucleons exceeding $\sim$ $\SI{5e19}{eV}$ (GZK process). Also, photons are produced after pair production for nucleons of EeV energy and above ($\SI{1}{EeV} = \SI{e18}{eV}$).
This is, in principle, a guaranteed production channel of UHE photons. However, the lack of knowledge about the initial flux of nucleons above
the GZK threshold translates into a substantial uncertainty when quantifying the expected production rate of cosmogenic photons. 
In addition, once created, UHE photons can initiate electromagnetic cascades, with an attenuation length of $\sim$$\SI{10}{Mpc}$ above $\SI{e19}{eV}$, shrinking to $\sim$$\SI{0.1}{Mpc}$ towards $\SI{e17}{eV}$~\cite{Heiter:2017cev}.
Thus, in case of distant sources, the energies of the cascade particles might fall below the detection threshold of air-shower observatories.

In the literature, there are only few works with predictions on the absolute flux of cosmogenic photons that experimental limits can be compared to.
In~\cite{Gelmini:2022evy}, while exploring the possibility to obtain an upper limit on the extragalactic radio background in case of photon observations,
photon fluxes were obtained for proton-emitting sources.
Photon fluxes assuming proton and iron primaries were determined in~\cite{Sarkar:2011hkm}.
For specific choices of model parameters, photon fluxes were derived in~\cite{Muzio:2019leu} when studying a sub-dominant UHE proton component allowed by cosmic-ray and neutrino data, and in~\cite{Alhebsi:2026bdk} when inspecting implications of a cosmogenic origin of the neutrino event KM3-230213A.

The intention of this work is to expand previous flux predictions. 
First, we investigate in detail the dependence of the photon yield on various parameters. Then, we provide updated simulations of the cosmogenic photon flux as well as a set of benchmark fluxes that reflects the range of uncertainty in the predictions.
We choose source scenarios that were identified to provide a good description of the present data from the Pierre Auger Observatory.
This will lead to a range of ``realistic'' flux predictions on cosmogenic photons.
In addition, we investigate the boundary flux levels (very high and very low fluxes) that emerge when relaxing some of the conditions for the allowed source scenarios.

In Sec.~\ref{sec:sims}, the simulation setup is described.
In Sec.~\ref{sec:fixed}, photon yields for sources at fixed energy and distance are studied. This will help to understand the relation between key source parameters and the photon flux at Earth.
In Sec.~\ref{sec:continuous}, photon spectra in scenarios with continuous injection energies and source distributions are regarded, varying parameters such as spectral index, cutoff energy, and primary mass.
In Sec.~\ref{sec:scenarios}, the cosmogenic photon flux is determined for a selection of six source scenarios.
A discussion of the results and a conclusion is given in Sec.~\ref{sec:conclusion}.


\section{Simulation setup}
\label{sec:sims}

The simulations are conducted with CRPropa 3.2~\cite{AlvesBatista:2022vem} in the one-dimensional mode. Within the CRPropa code, all interactions relevant for the present study are taken into account. For protons, neutrons, and nuclei, this includes photo-pion production, pair production and (for nuclei) photo-disintegration and elastic scattering in interactions with background photons. Decay of radioactive nuclei is considered as well. For secondary photons and electrons, the relevant electromagnetic processes include pair production, double and triplet pair production, inverse Compton scattering and, in the presence of magnetic fields, synchrotron emission.
For a detailed description of all processes relevant for the production and propagation of UHE photons, we refer to~\cite{Heiter:2017cev}.

Unless for studies with fixed distances, we consider scenarios with homogeneously distributed sources with a minimum distance of $\SI{1}{Mpc}$ and a maximum distance of $\SI{1000}{Mpc}$, as contributions to the UHE photon flux from even larger distances were found to be comparably small. If not stated otherwise, the photon background fields from~\cite{Gilmore:2011ks} and~\cite{Protheroe:1996si} are used, no magnetic fields have been switched on, and no source evolution is assumed. The typical number of simulated events per setting varies depending on the specific study. For initial studies with fixed initial energies and distance, a minimum of $10^5$ initial particles was considered per combination. For studies with continuous energies and distances, at least $4.2 \times 10^6$ primaries were simulated for each setting with energies above $10^{17.8}$~eV. 

To obtain an absolute flux, the simulations are normalized to the cosmic-ray flux at $10^{19}$~eV as measured by the Pierre Auger Observatory~\cite{PierreAuger:2021hun}.


\section{Photon yields for fixed energy and distance}
\label{sec:fixed}

\begin{figure}
    \centering
    \includegraphics[width=\linewidth]{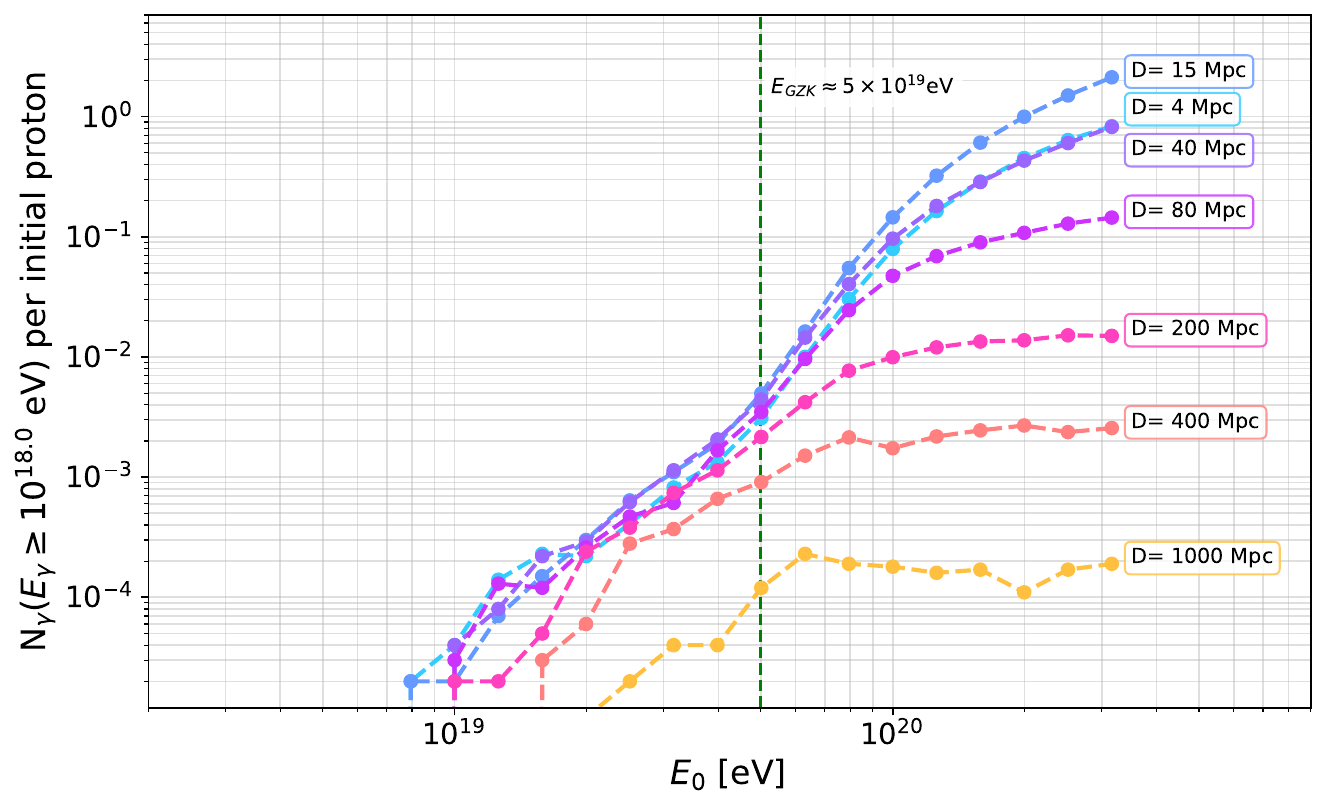}
    \caption{Number of photons at Earth above $10^{18}$~eV from initial protons depending on initial energy $E_0$ and source distance. The onset of the GZK process is indicated. For very low photon yields, the effect of statistical fluctuations becomes visible.}
    \label{fig:photon_yield_H}
\end{figure}

First, the dependence of the photon yield (number of photons at Earth with energies above a threshold of $10^{18}$~eV) on the initial energy, particle type and source distance is studied. 
For primary protons, the photon yield is shown in Fig.~\ref{fig:photon_yield_H} for different combinations of initial energy and distance.

The maximum photon yield is reached for a distance of about $\SI{15}{Mpc}$ due to a competition between UHE photon production and dilution.
The photon yield increases from $\sim$$10^{-4}$ at around $10^{19}$~eV to a remarkable value of about 1$-$2 UHE photons per initial proton at $(2-3)\times 10^{20}$~eV.
This strong increase is related to the increase / onset of the interaction cross-sections with energy (pair production and, at $\sim$$\SI{5e19}{eV}$, photon-pion production~/~GZK~effect). The onset of the GZK process is indicated in the plot. The dependence of the photon yield on distance is fairly mild below the GZK threshold, and quite pronounced above.
The photon yield decreases for $\SI{4}{Mpc}$ distance: for a source too close, the fraction of interacting protons gets reduced.
In turn, for large distances, the produced photons get more diluted by cascading to lower energy. Related to this, the maximum photon yield is mildly shifted from $\SI{15}{Mpc}$ to larger (smaller) distances when decreasing (increasing) the photon threshold energy.

\begin{figure}
    \centering
    \includegraphics[width=\linewidth]{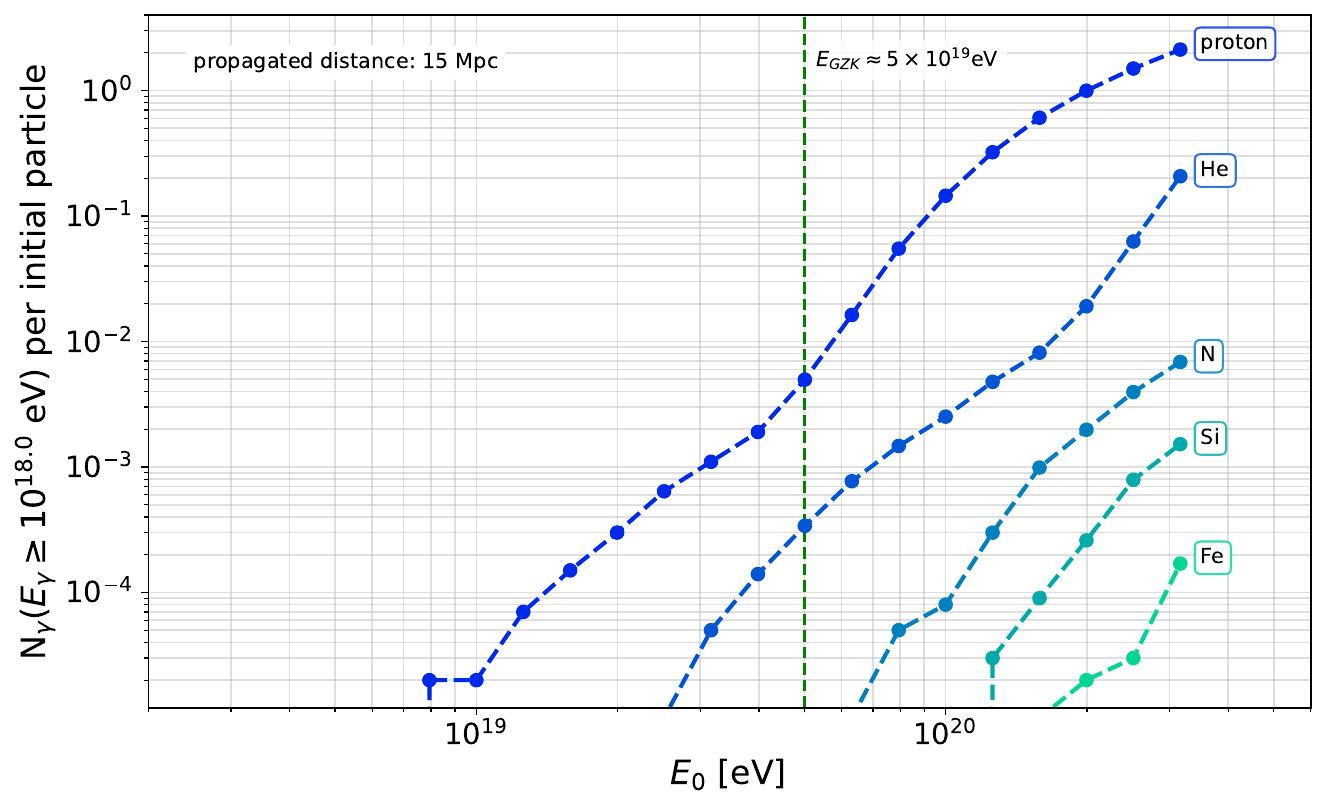}
    \caption{Number of photons at Earth above $10^{18}$~eV for different nuclei propagated over a distance of $\SI{15}{Mpc}$ against the initial energy $E_0$.}
    \label{fig:photon_yield_mixed}
\end{figure}

The photon yield for different nuclei is shown in Fig.~\ref{fig:photon_yield_mixed} for a fixed distance of $\SI{15}{Mpc}$. The significant decrease with increasing mass number $A$ can -- very approximatively -- be understood within the superposition model. Based on the photon yield for protons, which shows an average increase with energy by $N_\gamma^p (E) \propto E^{x}$ with $x\simeq 3.3$ in the relevant energy range, the photon yield for a nucleus of mass $A$ is estimated as $N_\gamma^A (E) \simeq A N_\gamma^p (E/A) \simeq N_\gamma^p (E) / A^{2.3}$, in reasonable agreement with the simulations.

\begin{figure}
    \centering
    \includegraphics[width=\linewidth]{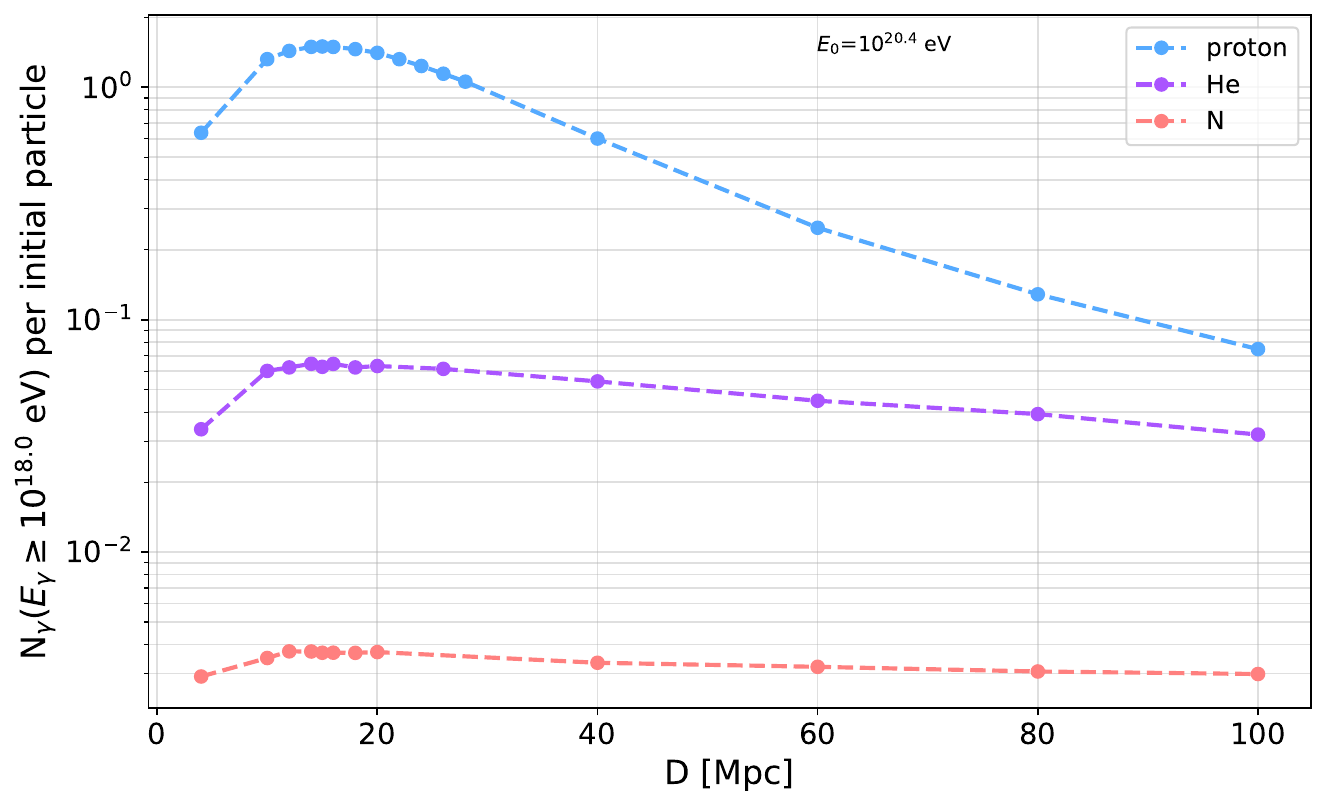}
    \caption{Number of photons at Earth above $10^{18}$~eV for different nuclei with initial energy of $10^{20.4}$~eV against the propagated distance.}
    \label{fig:photon_yield_dist_mixed}
\end{figure}

In Fig.~\ref{fig:photon_yield_dist_mixed}, we show the photon yield as a function of distance for different primaries with an energy of $10^{20.4}$~eV. For very close sources, the suppression of the yield is visible also for nuclei. With growing distance, however, the yield changes much more strongly for protons. Between $\SI{15}{Mpc}$ and $\SI{100}{Mpc}$, the yield decreases by a factor $\sim$25 for protons, but only by a factor $\sim$2 for helium, and even less for heavier nuclei. As a consequence, the yields from proton and helium are approaching each other for distant sources, differing by only a factor $\sim$2.3 at $\SI{100}{Mpc}$.

\begin{figure}
    \centering
    \includegraphics[width=\linewidth]{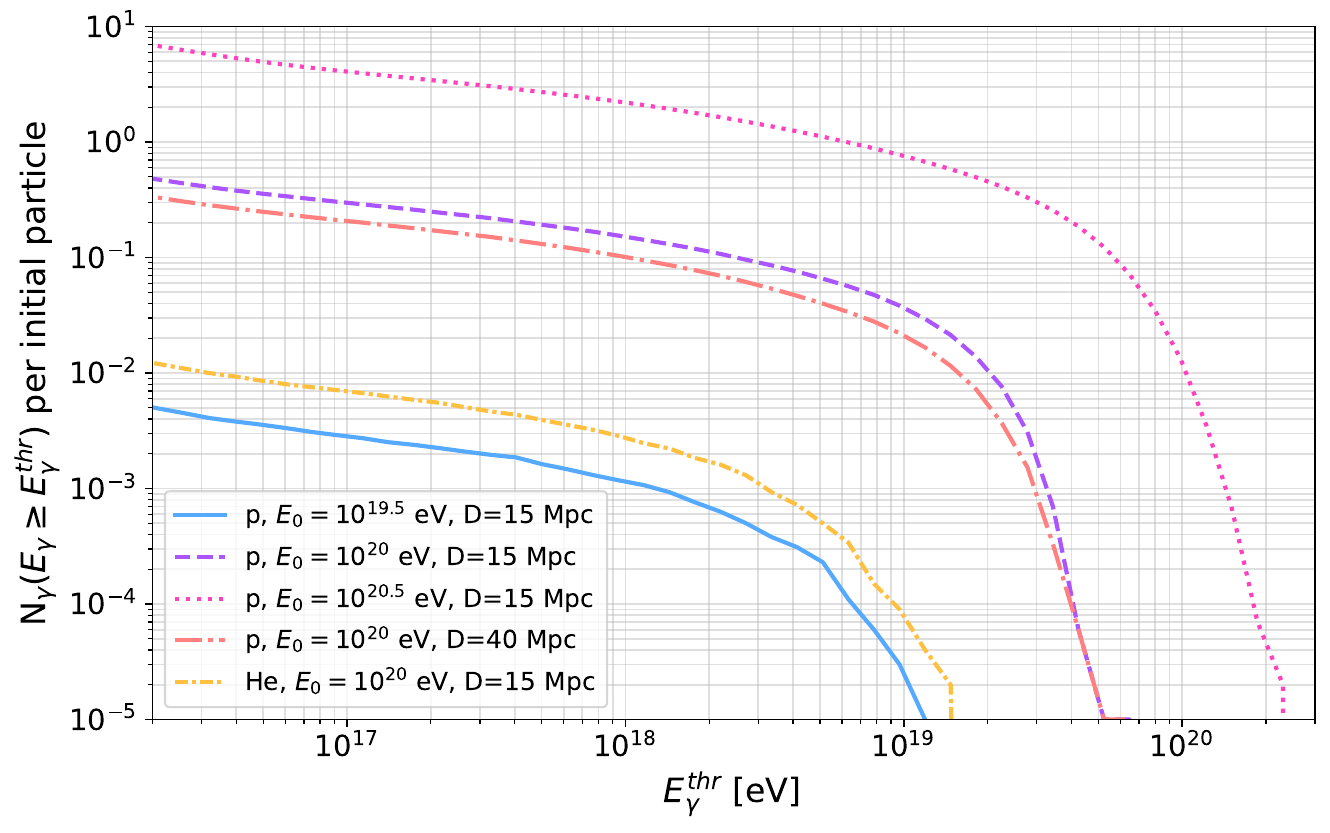}
    \caption{Number of photons at Earth as a function of the photon threshold energy $E_\gamma^{\text{thr}}$ for different initial energies, primaries, and propagation distances.}
    \label{fig:photon_spec_from_fix_E}
\end{figure}

So far, the studies of the photon yield focused on a fixed photon threshold of $E_\gamma^{\text{thr}} = 10^{18}$~eV. In Fig.~\ref{fig:photon_spec_from_fix_E}, the integral photon yield at Earth is shown as a function of $E_\gamma^{\text{thr}}$.
The photon spectrum is shown for $10^{20}$~eV protons from $\SI{15}{Mpc}$ as well as for variations of these parameters in terms of energy, distance and primary type. Overall, the spectral shapes are quite similar unless approaching the initial energy per nucleon. The flux decreases by about a factor $\sim$4 between $10^{18}$~eV and $10^{19}$~eV for initial energies above the GZK threshold, with an accelerated decrease when the photon threshold energy reaches 10$-$20\% of the initial energy per nucleon.

\section{Photon spectra for continuous energies and distances}
\label{sec:continuous}

\begin{figure}
    \centering
    \includegraphics[width=\linewidth]{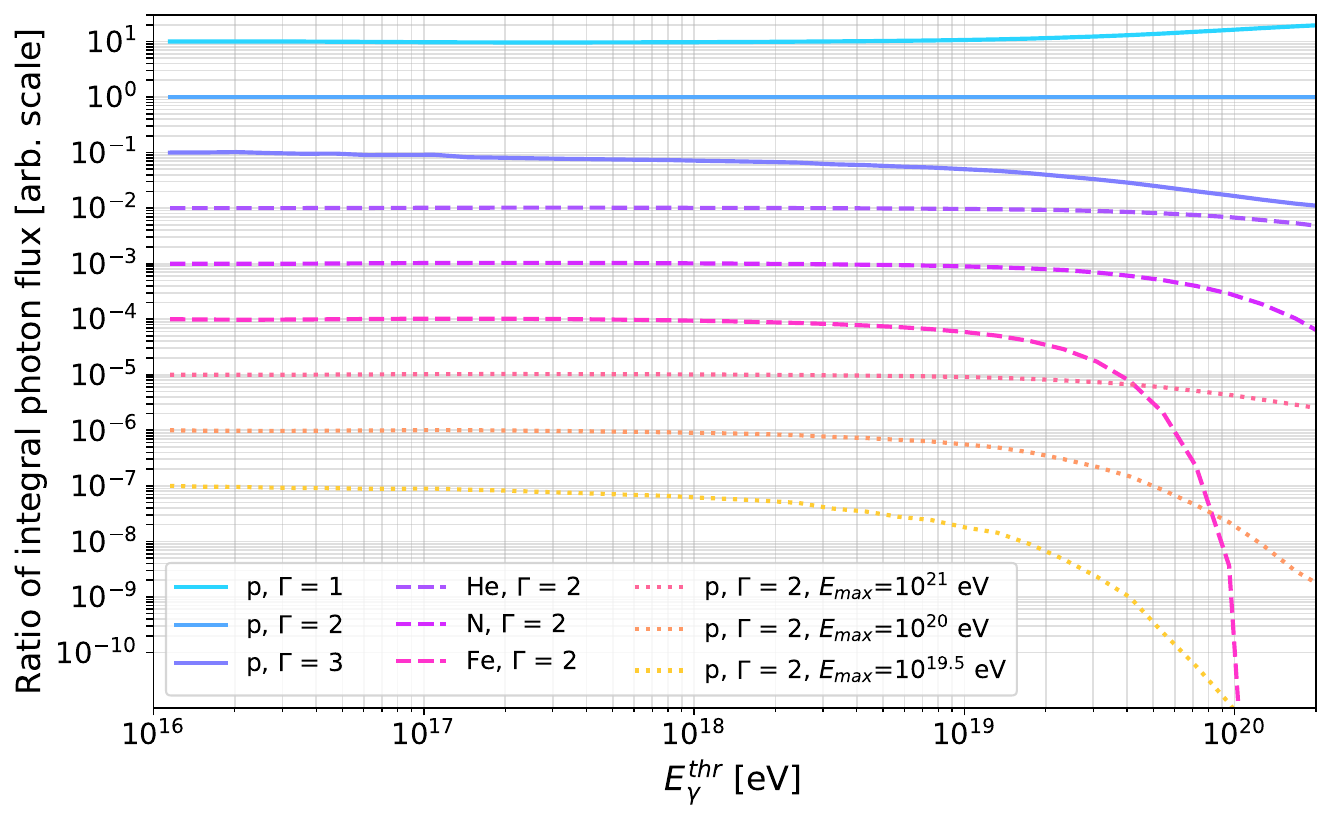}
    \caption{Shapes of photon spectra: integral photon flux vs.~photon energy threshold $E_\gamma^{\text{thr}}$ for protons and heavier nuclei from homogeneously distributed sources with power-law spectra for different values of $\Gamma$ and different cutoff energies $E_\text{max}$ as indicated. The spectra are divided by the curve for initial protons with $\Gamma=2$, and vertically shifted for better visibility.}
    \label{fig:spectral_shapes}
\end{figure}

To inspect the influence of spectral parameters and approach more realistic scenarios, simulations have been performed for initial energy spectra of the form $\Phi\propto E^{-\Gamma}$, both with and without cutoffs, from sources distributed uniformly over a comoving distance range.
Results are shown in Fig.~\ref{fig:spectral_shapes} for various combinations of initial spectral parameters and injected particles. The focus here is on the relative shapes of the different curves of the resulting photon spectra rather than on the absolute scale since the absolute flux scale is somewhat arbitrary for settings that do not reproduce the cosmic-ray flux well. Thus, the curves are divided by the case of initial protons with spectral index $\Gamma = 2$, and vertically shifted for better visibility.

Comparing spectra for initial protons for different values of $\Gamma$ (solid lines in Fig.~\ref{fig:spectral_shapes}), the shapes are quite similar for energy thresholds below $10^{18}$~eV. For higher $E_\gamma^{\text{thr}}$, steeper spectra lead to an accelerated reduction of the flux. For instance, comparing the case of $\Gamma=3$ to $\Gamma=2$, an additional suppression by a factor $\sim$5 is seen when moving the photon threshold from $10^{18}$~eV to $10^{20}$~eV. This is understandable as the relative proportion of high-energy initial protons is reduced.
The effect of accelerated reduction with increased photon threshold is also visible for initial nuclei (dashed lines in Fig.~\ref{fig:spectral_shapes}) where the fraction of particles with an energy per nucleon above the GZK threshold is (much) lower compared to the proton case. As an example, for $\Gamma = 2$ there is an additional suppression for initial iron by an order of magnitude compared to the proton case for $E_\gamma^{\text{thr}} = \SI{4e19}{eV}$, followed by a steep suppression towards higher energies.

Next, high-energy cutoffs are introduced (dotted lines in Fig.~\ref{fig:spectral_shapes}). The power-law spectrum is modified here by a simple exponential, i.e., $\Phi\propto E^{-\Gamma}\times f_\text{cut}$ with

\begin{equation}
\label{eq:cutoff_simple}
f_\text{cut} (E, E_\text{max}) = e^{-E/E_{\text{max}}}
\end{equation}

Results are shown for initial protons with $\Gamma=2$
for different values of $E_\text{max}$.
The impact of $E_\text{max}= 10^{21}$~eV is still moderate for photon energy thresholds below $10^{20}$~eV.
For $E_\text{max}=$ $10^{20}$~eV, the impact remains small for photon energy thresholds up to a few EeV. Increasing $E_\gamma^{\text{thr}}$ further towards $E_\text{max}$, the suppression of the initial protons becomes visible in the photon yield. Between photon thresholds of $10^{19}$~eV and $10^{20}$~eV, the photon flux is reduced by more than an order of magnitude compared to the case without cutoff. For $E_\text{max}= 10^{19.5}$~eV  $-$ corresponding to a flux suppression already below the GZK threshold $-$ the spectral shape is moderately affected up to photon thresholds below $10^{18}$~eV, followed by a reduction by a factor $\sim$5 at $E_\gamma^{\text{thr}} = 10^{19}$~eV, and a steep suppression further on.

\section{Cosmogenic photon fluxes for different source scenarios}
\label{sec:scenarios}

As is clear, the level of the cosmogenic photon flux may vary significantly, even by orders of magnitude, depending on the assumptions on the source parameters.
Of particular importance is the injection of nucleons with energies above the GZK threshold.
We therefore regard different source scenarios with the aim to obtain, first, a range of photon fluxes that appear realistic, given the present understanding of air-shower data.
And second, we want to provide benchmark flux levels for relaxed assumptions to indicate a very high (``maximum'') and very low (''minimum'') cosmogenic photon flux.

\begin{table}[]
            \begin{tabular}{|c|c|c|c|c|c|c|c|}
        \hline
               &  proton max& proton II & mixed Ia & mixed Ib& \multicolumn{2}{c|}{mixed II}  & iron min\\
                &  &  & & & LE & HE & \\
             \hline
             $\Gamma$ & $2.3$ & $2.25$ & $0.96$ & $2.04$ & $3.52$ & $-1.99$ & $2.5$ \\
             $\lg (E_\text{max}^p$/eV) & $21.0$ &$19.75$ &$18.68$ &  $19.88$  & $>$$19.4$ & $18.15$ & $19.0$ \\
             cutoff (Eq.) & (\ref{eq:cutoff_simple}) & (\ref{eq:cutoff_simple}) & (\ref{eq:cutoff_broken})  & (\ref{eq:cutoff_broken})& \multicolumn{2}{c|}{(\ref{eq:cutoff_broken})} & (\ref{eq:cutoff_simple}) \\
             $f_H$ & $1$ & $1$ & $0$& $0$ &$0.487$ & $0$ &  $0$\\
             $f_{He}$  & $0$ & $0$ & $0.673$& $0$ & $0.073$ & $0.236$ & $0$\\
             $f_N$    & $0$ & $0$ & $0.281$& $0.798$ & $0.44$ & $0.721$ & $0$\\
             $f_{Si}$  & $0$ & $0$ & $0.046$& $0.202$ & $0$ & $0.013$ & $0$\\
             $f_{Fe}$  & $0$ & $0$ & $0$& $0$ & $0$ & $0.031$ & $1$\\
              $m$   &$0$ & $5$ &$0$ &$0$ & $0$ & $0$ & $0$  \\
            \hline 
        \end{tabular}
    \caption{List of parameters in the six regarded scenarios (``proton II'' from~\cite{boncioli2025newviewuhecrspierre}, ``mixed Ia/b'' from~\cite{PierreAuger:2016use}, ``mixed II'' from~\cite{PierreAuger:2022atd}): spectral index $\Gamma$, cutoff energy $E_\text{max}^p$, cutoff type (with reference to Equation), particle fractions $f_i$ of nucleus $i$, source evolution parameter $m$. The ``mixed II'' scenario contains two components (LE and HE) with relative weighting $\text{LE}/\text{HE} \simeq 2.24/1$. For nuclei of charge number $Z$, it follows $E_\text{max} = Z E_\text{max}^p$.
}
    \label{tab:scenario_collection_parameters}
\end{table}

In total, we choose six scenarios, the details of which are listed in Tab.~\ref{tab:scenario_collection_parameters}.
Three scenarios are based on results obtained from fitting the energy spectrum and mass composition measured by the Pierre Auger Observatory,
leading to mixed compositions at injection with low or even zero proton components.
Two of these scenarios (``mixed Ia'' and ``mixed Ib'') are from~\cite{PierreAuger:2016use}, one scenario (''mixed II'') is from~\cite{PierreAuger:2022atd}. Other scenarios from~\cite{PierreAuger:2016use,PierreAuger:2022atd}
lead to photon fluxes bracketed by these three scenarios.
The cutoff term is here described by
\begin{align} 
\label{eq:cutoff_broken}
f_\text{cut}(E, E_\text{max})=\begin{cases} e^{1-E/E_{max}} &\text{if } E> E_\text{max},   \\ 
    1 &\text{else},\end{cases} 
\end{align}
where $E_\text{max} = Z E_\text{max}^p$
scales with the charge number $Z$ of a nucleus compared to the cutoff energy $E_\text{max}^p$ of protons. 
The ``mixed II'' scenario consists of two components (LE and HE). The cutoff energy $E_\text{max}^p$ is degenerate in the LE component, allowing for any value above $10^{19.4}$~eV. Thus, we inspect the two limiting cases of $E_\text{max,LE}^p = 10^{19.4}$~eV and $E_\text{max, LE}^p \rightarrow \infty$ within the ``mixed II'' scenario.

Next, a scenario discussed in~\cite{boncioli2025newviewuhecrspierre} is chosen. This is a pure proton scenario (termed here ``proton II'') with injection parameters such that it leads to a reasonable description of the Auger spectrum at the highest energies. The cutoff function is in this case given by Eq.~(\ref{eq:cutoff_simple})~\cite{Denise_Boncioli}.
As noted in~\cite{boncioli2025newviewuhecrspierre}, for a large evolution parameter $m$, this scenario might overshoot the Auger limits on the neutrino flux, and details of the EeV cosmic-ray flux might be difficult to reconcile.
Moreover, possible constraints may arise from  gamma-ray limits at GeV$-$TeV energies~(e.g.,~\cite{Fermi-LAT:2014ryh,Berezinsky:2016feh}).
And, of course, composition data are not used. 
Thus, this scenario provides an example of a photon flux from initial protons when focusing only on the cosmic-ray flux at highest energies.

Finally, two extreme scenarios are regarded in which we try to maximize (minimize) the photon flux with the only boundary condition that the resulting cosmic-ray spectrum above $10^{19}$~eV is not contradicting the data beyond the present levels of uncertainty.
The maximized photon flux (scenario ``proton max'') is then obtained using pure protons with a fairly large cutoff energy and a moderate spectral index,
and the minimized photon flux (scenario ``iron min'') in turn for pure iron with fairly small cutoff energy.
A moderate rescaling (of 20$-$30\%) of the overall flux has been applied in both of these two scenarios so that the simulated cosmic-ray flux scratches the upper band (for ``proton max'') or lower band (for ``iron min'') of the measured cosmic-ray spectrum at the highest energies, since it is the highest energies that dominate the resulting photon yield.
The proton scenario is similar to the scenario presented 
in~\cite{Sarkar:2011hkm} but has a steeper spectrum to better fit the observed suppression of the high-energy cosmic-ray flux.

We note that photon fluxes from the pure proton scenarios might well be regarded as (too) optimistic, but might be useful for more complex scenarios
where the proton component contributes only a fraction of the injected cosmic rays (e.g., when accounting for composition and multimessenger constraints). Then, the proton contribution to the cosmogenic photon flux can be scaled accordingly.

\begin{figure}
    \centering
    \includegraphics[width=\linewidth]{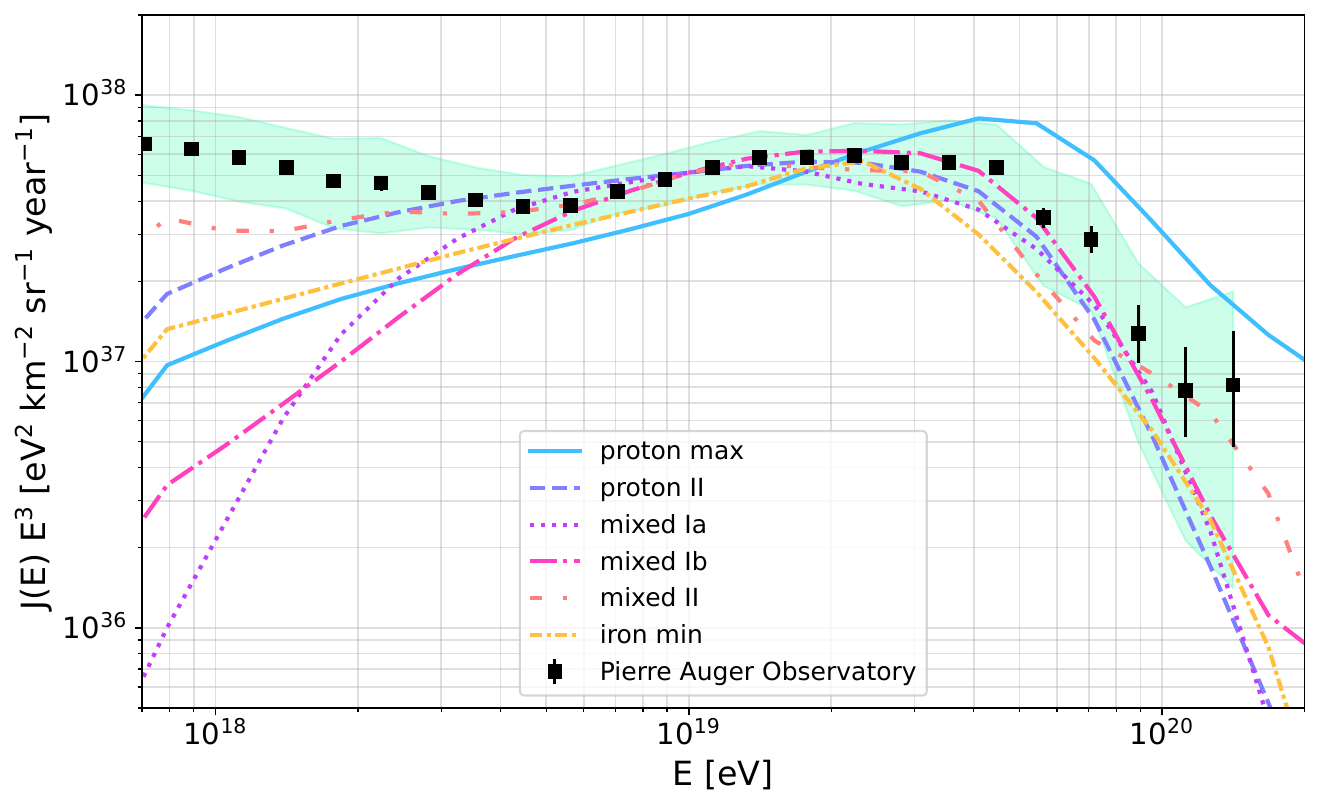}
    \caption{Cosmic-ray fluxes of the six considered source scenarios and the flux measured by the Pierre Auger Observatory~\cite{PierreAuger:2021hun}. The error bar indicates the statistical and the band the added statistical and systematic uncertainty.}
    \label{fig:scenario_collection_CR}
\end{figure}

\begin{figure}
    \centering
    \includegraphics[width=\linewidth]{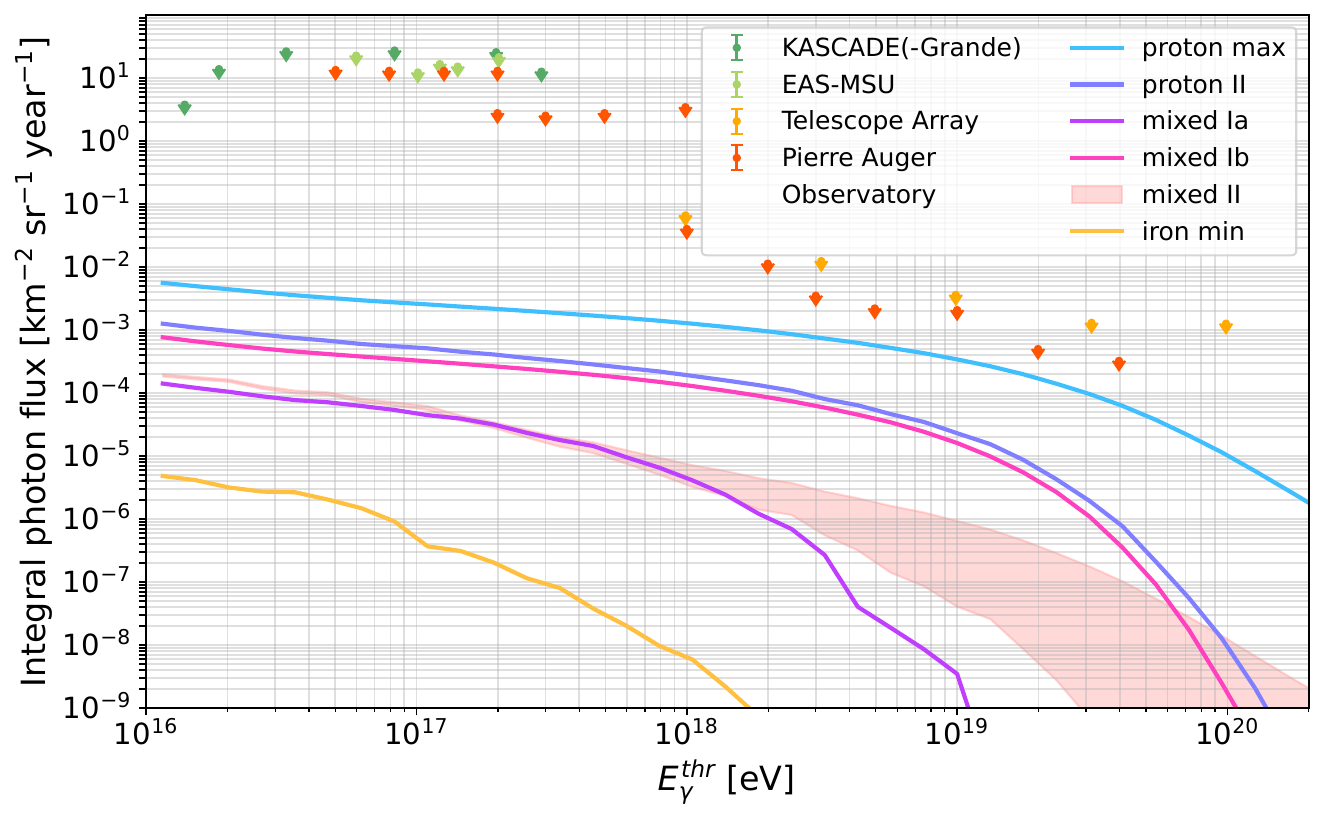}
    \caption{Integral flux of cosmogenic photons as a function of photon energy threshold $E_\gamma^{\text{thr}}$ of the six considered source scenarios. Also shown the present experimental upper limits on photons (taken from~\cite{PierreAuger:2025jwt}).}
    \label{fig:scenario_collection_photon}
\end{figure}

The results for all six scenarios are shown in Fig.~\ref{fig:scenario_collection_CR} (cosmic-ray spectrum) and Fig.~\ref{fig:scenario_collection_photon} (integral photon flux). Also shown in Fig.~\ref{fig:scenario_collection_photon} are the present experimental upper limits on photons taken from~\cite{PierreAuger:2025jwt}.
We remind the reader that these scenarios are not supposed to fit the cosmic-ray flux at energies below $(5-10)\times 10^{18}$~eV. Additional components would be needed, e.g., from a new source class or, partially, from contributions of sources beyond the simulation limit of 1000~Mpc. In principle, these components might add to the photon flux below a few EeV. However, since the UHE photon flux is dominated by the high-energy part of cosmic rays, this contribution is expected to be comparably small. The two limiting cases of $E_\text{max,LE}^p$ regarded within the ``mixed II'' scenario do not have a visible impact on the cosmic-ray flux (in line with the findings of~\cite{PierreAuger:2022atd}) such that only one curve is given in Fig.~\ref{fig:scenario_collection_CR} for this scenario.

Turning to the results on cosmogenic photons in Fig.~\ref{fig:scenario_collection_photon}, the highest photon flux is obtained, by construction, in scenario ``proton max''.
Above $10^{18}$~eV ($10^{19}$~eV), the flux amounts here to $\sim$$1.3 \cdot 10^{-3}$~km$^{-2}$~sr$^{-1}$~yr$^{-2}$ ($3.4 \cdot 10^{-4}$~km$^{-2}$~sr$^{-1}$~yr$^{-2}$). While not excluded by present experimental limits,
testing such a scenario might be in reach (see also Sec.~\ref{sec:conclusion}).

The flux level of scenario ``proton II'' is reduced, compared to ``proton max'', by about an order of magnitude for photons above $10^{18}$~eV. This is primarily due to the smaller value of $E_{\text{max}}$. The reduction is smaller (larger) at lower (higher) threshold energy,
in line with the results from Sec.~\ref{sec:continuous}.

Interestingly, scenario ``mixed Ib'' leads to a level of cosmogenic photons that is similar to the one of ``proton II'' within a factor $\sim$2, varying only slightly depending on the threshold energy.
This scenario does not contain initial protons or helium. The dominant contribution comes from nitrogen. As the value of $E_\text{max}^p$, however, allows for nucleons above the GZK energy threshold, the production of GZK photons is possible, leading to a comparably high photon flux.

The flux levels of scenarios ``mixed Ia'' and ``mixed II'' are similar to each other up to threshold energies of $\sim$$10^{18}$~eV, and about one order of magnitude below ``mixed Ib''.
At higher $E_\gamma^{\text{thr}}$, ``mixed II'' exceeds ``mixed Ia'' due to a larger population of highest-energy cosmic rays.
Also, an impact of $E_\text{max,LE}^p$ becomes visible (indicated by the band in Fig.~\ref{fig:scenario_collection_photon}): when effectively removing the cutoff, photons above $10^{19}$~eV appear, related to the injection of protons within the LE component. The photon flux even reaches the levels of the ``mixed Ib'' and ``proton II'' scenarios at very high photon energies. In turn, adopting $E_\text{max,LE}^p = 10^{19.4}$~eV, the high-energy photon flux is strongly reduced, and even more so in ``mixed Ia'', due to the
suppressed injection of nucleons above the GZK threshold.
Thus, it is interesting to see that the degeneracy of $E_\text{max,LE}^p$ (cutoff energy of LE component in the ``mixed II'' scenario) is broken by cosmogenic photons: while leaving the overall cosmic-ray flux unchanged, the photon flux varies by more than 1.5 orders of magnitude for photon energies above $10^{19}$~eV, depending on the parameter $E_\text{max,LE}^p$.

A further strong decrease of the photon flux is obtained in the low-flux benchmark ``iron min''.
Compared to the ``proton max'' scenario, the flux is reduced by more than three (five) orders of magnitude at $10^{17}$~eV ($10^{18}$~eV), and much more at higher energies. While this scenario might anyway be regarded as (too) pessimistic, it is worthwhile to note that a floor of cosmogenic photons around and above $10^{18}$~eV is expected even here. Still, it is difficult to imagine experimental sensitivities coming close to testing such flux levels in the foreseeable future.

\section{Discussion and conclusion}
\label{sec:conclusion}

We checked the dependence of the photon flux on further parameters by varying the magnetic field strength, the radio and infrared background and the source evolution.

Allowing for a uniform, perpendicular magnetic field, the effect is very minor (few percent) for field strengths up to 0.1~nG.
Adopting a larger magnetic field, the photon yield gets more significantly reduced due to synchrotron emission becoming a competing process for electrons and positrons to the production of a secondary UHE photon by inverse Compton scattering.
For instance, for a 1~nG field, the photon yield (for $E_\gamma^{\text{thr}} = 10^{18}$~eV) of $10^{20}$~eV protons decreases by a factor $\sim$2.

The radio background affects the interaction length of photons in the cascade.
Switching from the radio background of Protheroe and Biermann~\cite{Protheroe:1996si} to that of Ni{\c{t}}u et al.~\cite{Nitu:2020vzn}, the interaction length is reduced, leading to a smaller UHE photon yield due to accelerated cascading.
As an example, for $10^{20}$~eV protons the photon yield decreases by a factor $\sim$1.5.
Similarly, the interaction length of photons is affected by the infrared background. By applying background models by Dominguez et al.~\cite{Dominguez:2010bv} and Stecker et al.~\cite{Stecker:2005qs}, the interaction length decreases and increases, respectively, compared to the model by Gilmore et al.~\cite{Gilmore:2011ks}. 
The photon yield from primary protons varies by less than $1\%$ between these models (cf.~also~\cite{Heiter:2017cev}). For primary iron nuclei at $\SI{3e20}{eV}$, the variation is below a factor of $1.5$.
For a detailed discussion of magnetic fields and photon backgrounds, we refer to~\cite{Gelmini:2022evy}.

We studied a possible evolution of the source injection power with redshift $z$ according to $(1+z)^m$.
The previously regarded scenarios contained values of the source evolution parameter of $m = 0$ and $m=5$.
As an example, for scenario ``proton II'', adopting $m = \pm3$ ($\pm6$) changes the flux of photons above $10^{18}$~eV by a factor $\sim$1.2 ($\sim$1.5) compared to the case of $m=0$. Here, one has to note that also the overall cosmic-ray flux is affected such that each scenario has individually been normalized to resemble the observed flux at $10^{19}$~eV.
The quoted impact on the photon flux when changing $m$ is reduced if the flux normalization is conducted at higher energy.

In~\cite{Gelmini:2022evy}, the effect of these parameters is also regarded.
While a direct comparison is difficult due to a different source scenario and calibration procedure used,
also in that work a minor effect for a 0.01 nG magnetic field and of the evolution parameter $m$ has been found.
And artificially switching off the
Protheroe and Biermann radio background increased the photon flux above $10^{18}$~eV by a factor $\sim$2, in line with the level of uncertainty we find related to the radio background.

Overall, while these parameters may (partially) have a visible impact on the cosmogenic photon flux, the uncertainty of the flux prediction is dominated by far by the uncertainties of the injection parameters as reflected in the different scenarios inspected in Sec.~\ref{sec:scenarios}.

\begin{figure}
    \centering
    \includegraphics[width=\linewidth]{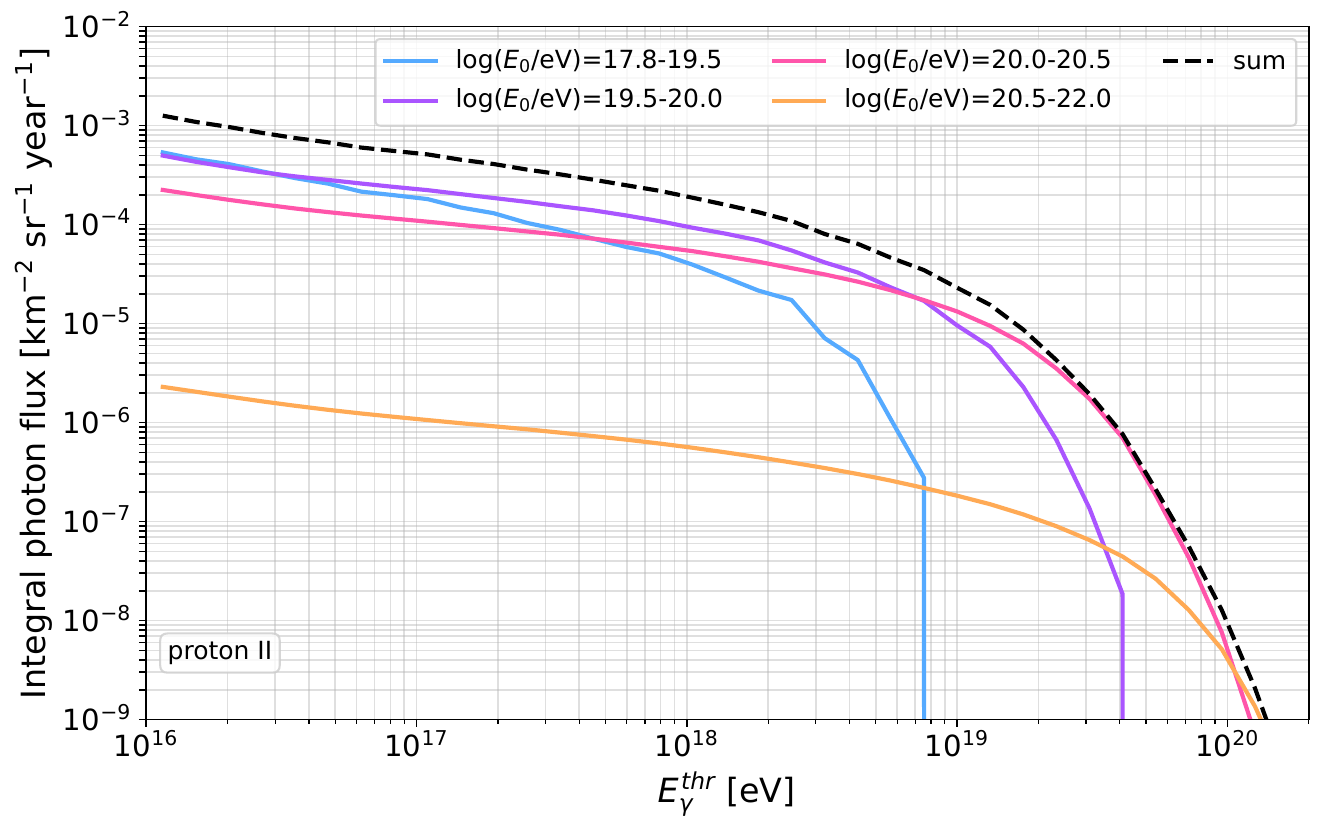}
    \caption{Contribution of different ranges of initial energies to the photon spectrum for the ``proton~II'' scenario.}
    \label{fig:protonII_energy_photon}
\end{figure}

\begin{figure}
    \centering
    \includegraphics[width=\linewidth]{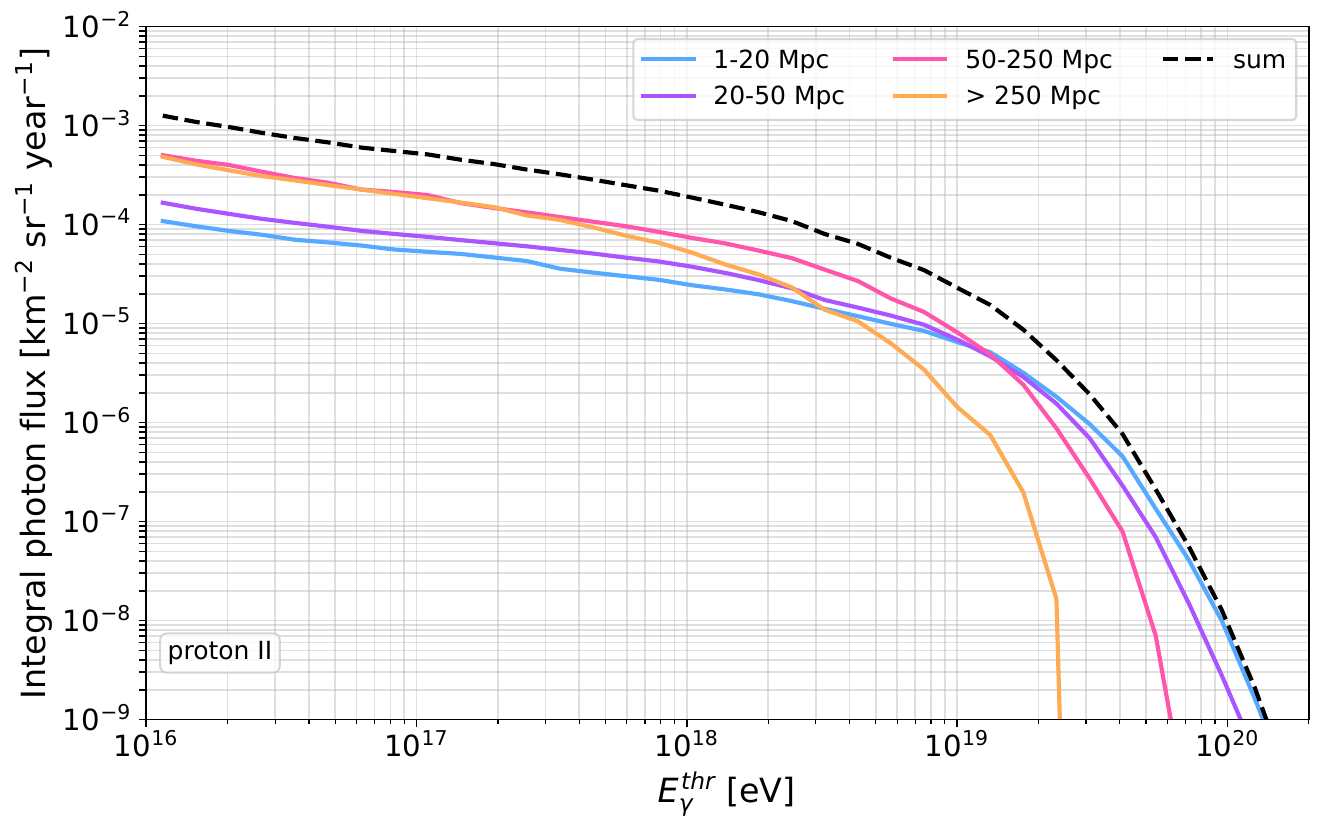}
    \caption{Contribution of different distance ranges to the photon spectrum for the ``proton~II'' scenario.}
    \label{fig:protonII_distance_photon}
\end{figure}

Within these scenarios, we analyzed which initial energies and which source distances contribute most to the resulting photon flux at Earth. For scenario ``proton II'' as an example, the contributions of different ranges of initial energies and of sources distances are plotted in Figs.~\ref{fig:protonII_energy_photon} and~\ref{fig:protonII_distance_photon}.
The photon flux is dominated by initial energies of
$\lg (E$/eV) = $[19.5,20.0]$ for photon thresholds up to $\sim$$\SI{7e18}{eV}$, and for higher thresholds by initial energies of $\lg (E$/eV) = $[20.0,20.5]$. Energies below $\lg (E$/eV) = 19.5 only show contributions for photons below $(1-2)\times 10^{18}$~eV, and energies above $\lg (E$/eV) = 20.5 only for photons above $\SI{4e19}{eV}$.
Concerning the source distances, it turns out that all regarded distance ranges contribute, with a clear dependence on photon threshold energy. Sources at smaller distances (ranges 1$-$20 and 20$-$50 Mpc) dominate for photons above $\sim$$\SI{1.3e19}{eV}$. Lowering $E_\gamma^{\text{thr}}$, the distance range 50$-$250 Mpc becomes important. Interestingly, also distances above 250 Mpc contribute for photons below $10^{19}$~eV, reaching a fraction of $\sim$35\% for $E_\gamma^{\text{thr}} = 10^{17}$~eV. For completeness, we mention that the contribution from sources beyond 1000~Mpc is found to be less than 3\% for photons above $10^{18}$~eV, shrinking quickly when increasing $E_\gamma^{\text{thr}}$. For photon thresholds below $10^{17}$~eV, the contribution increases to $\sim$10\%.

For the ``mixed Ib'' scenario, the contributing source distances are quite similar to those for ``proton II'', with a slightly reduced fraction from distant sources. In contrast, in this scenario the photon yield is dominated by far only by the highest initial energies above $\lg (E$/eV) = 20.5. This is understandable as only at these energies, the energy per nucleon exceeds the GZK threshold.
Also in the ``proton max'' scenario, the highest energy range dominates the photon flux. In this scenario, however, the relative contribution from closer sources is more pronounced, with source distances above 250 Mpc having only a minor impact.
For completeness, we mention that the differential plots for the contributing ranges tend to all have a similar shape.

We compared results on cosmogenic photons from our simulation setup to the previous ones obtained for pure proton and iron in~\cite{Gelmini:2022evy,Sarkar:2011hkm}, trying to reproduce those parameter choices. We find reasonable agreement, with differences below a factor $\sim$2. Details on a comparison to the work of~\cite{Sarkar:2011hkm} can be found in~\cite{Papior:2025xjd}. Such residual differences might be expected due to differences in the used propagation codes, the flux calibrations and details of the propagation parameters adopted. For instance, while~\cite{Sarkar:2011hkm} also used CRPropa, a major
upgrade of the code was conducted in the meantime concerning the simulation of photon interactions and propagation~\cite{Heiter:2017cev}. In~\cite{Gelmini:2022evy}, a custom numerical code was applied along with a tailor-made flux calibration using the highest-energy spectrum measured by the Telescope Array.
For the mixed composition scenarios, our results differ from those of an initial study~\cite{Bobrikova:2021kuj} due to a technical problem in that work.

\begin{figure}
    \centering
    \includegraphics[width=\linewidth]{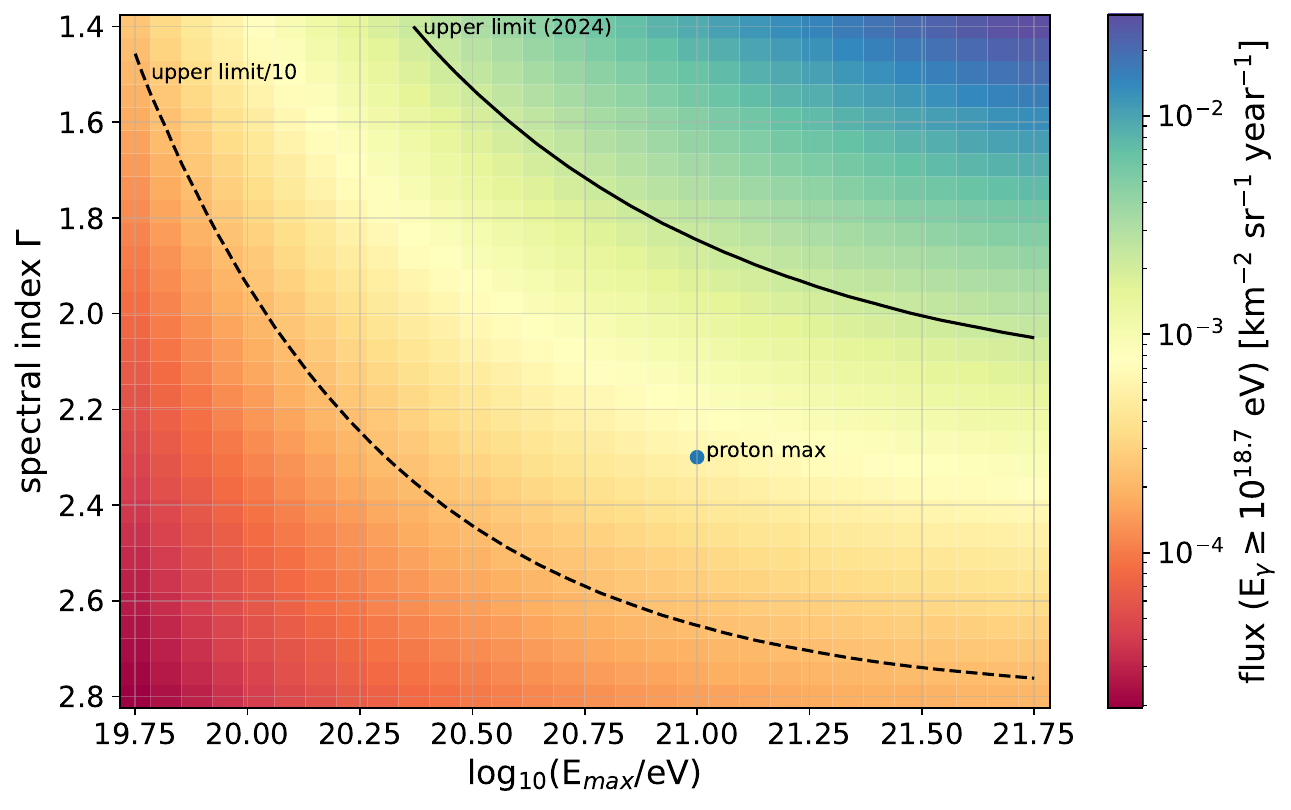}
    \caption{Integral photon fluxes above $10^{18.7}$~eV expected for pure-proton scenarios similar to the ``proton max'' scenario when varying $E_\text{max}$ (cf.~Eq.~(\ref{eq:cutoff_simple})) and spectral index $\Gamma$. Values for the ``proton max'' scenario are marked. Lines indicate the present upper limit~\cite{PierreAuger:2024ayl} and the flux level in case of a factor 10 improvement.}
    \label{fig:Rmax_gamma_scan}
\end{figure}

As the ``proton max'' scenario is closest to the experimental limits, we checked which combinations of $E_\text{max}$ and $\Gamma$ can be excluded in a pure proton scenario with present limits. The result is given in Fig.~\ref{fig:Rmax_gamma_scan} for a photon threshold of $10^{18.7}$~eV. Certain combinations of a comparably flat spectrum and a high cutoff can be excluded due to the non-observation of a related high flux of cosmogenic photons. Also indicated is the line for limits improved by an order of magnitude. For testing the ``proton max'' scenario, an improvement of a factor $\sim$3 is needed. With a continuously growing exposure and an improved photon-hadron separation, this sensitivity might be in reach for the recently upgraded Pierre Auger Observatory~\cite{PierreAuger:2025lsh}.

In summary, after investigating the dependence of the photon yield on various parameters, we provided the cosmogenic photon flux for a set of different scenarios. Defining two extreme benchmark scenarios by neglecting composition data, the flux level in the ``proton max'' scenario might be in reach. The flux level in the ``iron min'' scenario is extremely low, but non-zero at EeV energies. Thus, a floor of cosmogenic UHE photons is expected to exist.

For realistic scenarios taking composition data into account, flux levels are more than 1.5~orders of magnitude below present limits, posing a challenge for observations. As these scenarios contain a very low or even zero proton fraction, it is worthwhile to note that an additional, even partial contribution from a proton-emitting source could lead to an increased flux level. This holds in particular for a source located close to the distance of the maximum photon yield of about 15~Mpc (which happens to be close to the distance to, e.g., the Virgo cluster). Also, in case of a fairly close source, the arrival directions of cosmogenic photons could possibly be used to further improve the experimental sensitivity. 
Finally, we note that there is also a connection to UHE neutrinos: in some production channels, cosmogenic neutrinos are produced alongside cosmogenic photons. Experimental limits on the photon flux together with limits on the neutrino flux can be used to further constrain the composition and source parameters of UHE cosmic rays. This makes the study of the flux levels of cosmogenic photons a cornerstone of multimessenger astronomy at the highest energies.

\acknowledgments

This work was supported by the German Research Foundation (DFG Project No. 508269468). We thank our colleagues from the Pierre Auger Collaboration for fruitful discussions. We are grateful to the referee for helpful comments. The simulations presented in this work were performed in part on the OMNI computing cluster of the University of Siegen.

\bibliographystyle{JHEP}
\bibliography{Bib.bib}

@article{Risse:2007sd,
    author = "Risse, Markus and Homola, Piotr",
    title = "{Search for ultrahigh energy photons using air showers}",
    eprint = "astro-ph/0702632",
    archivePrefix = "arXiv",
    doi = "10.1142/S0217732307022864",
    journal = "Mod. Phys. Lett. A",
    volume = "22",
    pages = "749--766",
    year = "2007"
}

@article{PierreAuger:2022gkb,
    author = "Abreu, Pedro and others",
    collaboration = "Pierre Auger",
    title = "{Searches for Ultra-High-Energy Photons at the Pierre Auger Observatory}",
    eprint = "2210.12959",
    archivePrefix = "arXiv",
    primaryClass = "astro-ph.HE",
    reportNumber = "FERMILAB-PUB-22-810-AD-PPD-SCD-TD",
    doi = "10.3390/universe8110579",
    journal = "Universe",
    volume = "8",
    number = "11",
    pages = "579",
    year = "2022"
}

@article{Heiter:2017cev,
    author = "Heiter, Christopher and Kuempel, Daniel and Walz, David and Erdmann, Martin",
    title = "{Production and propagation of ultra-high energy photons using CRPropa 3}",
    eprint = "1710.11406",
    archivePrefix = "arXiv",
    primaryClass = "astro-ph.IM",
    doi = "10.1016/j.astropartphys.2018.05.003",
    journal = "Astropart. Phys.",
    volume = "102",
    pages = "39--50",
    year = "2018"
}

@article{Gelmini:2022evy,
    author = "Gelmini, Graciela B. and Kalashev, Oleg and Semikoz, Dmitri",
    title = "{Upper Limit on the Diffuse Radio Background from GZK Photon Observation}",
    eprint = "2206.00408",
    archivePrefix = "arXiv",
    primaryClass = "astro-ph.HE",
    doi = "10.3390/universe8080402",
    journal = "Universe",
    volume = "8",
    number = "8",
    pages = "402",
    year = "2022"
}

@inproceedings{Sarkar:2011hkm,
    author = "Sarkar, Biswajit and Kampert, Karl-Heinz and Kulbartz, Joerg",
    title = "{Ultra-High Energy Photon and Neutrino Fluxes in Realistic Astrophysical Scenarios}",
    booktitle = "{Proc. 32nd International Cosmic Ray Conference}",
    doi = "10.7529/ICRC2011/V02/1087",
    volume = "2",
    pages = "198",
    year = "2011"
}

@article{AlvesBatista:2022vem,
    author = "Alves Batista, Rafael and others",
    title = "{CRPropa 3.2 \textemdash{} an advanced framework for high-energy particle propagation in extragalactic and galactic spaces}",
    eprint = "2208.00107",
    archivePrefix = "arXiv",
    primaryClass = "astro-ph.HE",
    doi = "10.1088/1475-7516/2022/09/035",
    journal = "JCAP",
    volume = "09",
    pages = "035",
    year = "2022"
}

@article{Gilmore:2011ks,
    author = "Gilmore, R. C. and Somerville, R. S. and Primack, J. R. and Dominguez, A.",
    title = "{Semi-analytic modeling of the EBL and consequences for extragalactic gamma-ray spectra}",
    eprint = "1104.0671",
    archivePrefix = "arXiv",
    primaryClass = "astro-ph.CO",
    doi = "10.1111/j.1365-2966.2012.20841.x",
    journal = "Mon. Not. Roy. Astron. Soc.",
    volume = "422",
    pages = "3189",
    year = "2012"
}

@article{Protheroe:1996si,
    author = "Protheroe, R. J. and Biermann, P. L.",
    title = "{A New estimate of the extragalactic radio background and implications for ultrahigh-energy gamma-ray propagation}",
    eprint = "astro-ph/9605119",
    archivePrefix = "arXiv",
    reportNumber = "ADP-AT-96-1, MPIFR-BONN-676",
    doi = "10.1016/S0927-6505(96)00041-2",
    journal = "Astropart. Phys.",
    volume = "6",
    pages = "45--54",
    year = "1996",
    note = "[Erratum: Astropart.Phys. 7, 181 (1997)]"
}

@article{PierreAuger:2021hun,
    author = "Abreu, P. and others",
    collaboration = "Pierre Auger",
    title = "{The energy spectrum of cosmic rays beyond the turn-down around $10^{17}$\,eV as measured with the surface detector of the Pierre Auger Observatory}",
    eprint = "2109.13400",
    archivePrefix = "arXiv",
    primaryClass = "astro-ph.HE",
    reportNumber = "FERMILAB-PUB-21-474-AD-AE-SCD-TD",
    doi = "10.1140/epjc/s10052-021-09700-w",
    journal = "Eur. Phys. J. C",
    volume = "81",
    number = "11",
    pages = "966",
    year = "2021"
}

@article{boncioli2025newviewuhecrspierre,
      author="{D. Boncioli for the \textsc{Pierre Auger} Collaboration}",
      title="{A new view of UHECRs with the Pierre Auger Observatory}", 
    doi = "10.22323/1.484.0027",
    journal = "PoS",
    volume = "UHECR2024",
    pages = "027",
    year = "2025",
          eprint={2509.15862},
      archivePrefix={arXiv},
      primaryClass={astro-ph.HE},
}

@article{PierreAuger:2016use,
    author = "Aab, Alexander and others",
    collaboration = "Pierre Auger",
    title = "{Combined fit of spectrum and composition data as measured by the Pierre Auger Observatory}",
    eprint = "1612.07155",
    archivePrefix = "arXiv",
    primaryClass = "astro-ph.HE",
    reportNumber = "FERMILAB-PUB-16-618",
    doi = "10.1088/1475-7516/2017/04/038",
    journal = "JCAP",
    volume = "04",
    pages = "038",
    year = "2017",
    note = "[Erratum: JCAP 03, E02 (2018)]"
}

@article{PierreAuger:2022atd,
    author = "Abdul Halim, A. and others",
    collaboration = "Pierre Auger",
    title = "{Constraining the sources of ultra-high-energy cosmic rays across and above the ankle with the spectrum and composition data measured at the Pierre Auger Observatory}",
    eprint = "2211.02857",
    archivePrefix = "arXiv",
    primaryClass = "astro-ph.HE",
    reportNumber = "FERMILAB-PUB-22-876-AD-PPD-SCD-TD",
    doi = "10.1088/1475-7516/2023/05/024",
    journal = "JCAP",
    volume = "05",
    pages = "024",
    year = "2023"
}

@article{PierreAuger:2025jwt,
    author = "Abdul Halim, Adila and others",
    collaboration = "Pierre Auger",
    title = "{Search for a diffuse flux of photons with energies above tens of PeV at the Pierre Auger Observatory}",
    eprint = "2502.02381",
    archivePrefix = "arXiv",
    primaryClass = "astro-ph.HE",
    doi = "10.1088/1475-7516/2025/05/061",
    journal = "JCAP",
    volume = "05",
    pages = "061",
    year = "2025"
}

@article{Nitu:2020vzn,
    author = "Ni{\c{t}}u, I. C. and Bevins, H. T. J. and Bray, J. D. and Scaife, A. M. M.",
    title = "{An updated estimate of the cosmic radio background and implications for ultra-high-energy photon propagation}",
    eprint = "2004.13596",
    archivePrefix = "arXiv",
    primaryClass = "astro-ph.HE",
    doi = "10.1016/j.astropartphys.2020.102532",
    journal = "Astropart. Phys.",
    volume = "126",
    pages = "102532",
    year = "2021"
}

@article{Papior:2025xjd,
    author = "Papior, Chiara Jane and Niechciol, Marcus and Risse, Markus",
    title = "{Estimating the GZK Photon Flux from Extragalactic Cosmic Rays}",
    doi = "10.22323/1.501.0353",
    journal = "PoS",
    volume = "ICRC2025",
    pages = "353",
    year = "2025"
}

@article{Bobrikova:2021kuj,
    author = "Bobrikova, Anna and Niechciol, Marcus and Risse, Markus and Ruehl, Philip",
    title = "{Predicting the UHE photon flux from GZK-interactions of hadronic cosmic rays using CRPropa 3}",
    doi = "10.22323/1.395.0449",
    journal = "PoS",
    volume = "ICRC2021",
    pages = "449",
    year = "2021"
}

@article{PierreAuger:2024ayl,
    author = "Abdul Halim, A. and others",
    collaboration = "Pierre Auger",
    title = "{Search for photons above $10^{18}$\,eV by simultaneously measuring the atmospheric depth and the muon content of air showers at the Pierre Auger Observatory}",
    eprint = "2406.07439",
    archivePrefix = "arXiv",
    primaryClass = "astro-ph.HE",
    reportNumber = "FERMILAB-PUB-24-0391-PPD-TD",
    doi = "10.1103/PhysRevD.110.062005",
    journal = "Phys. Rev. D",
    volume = "110",
    number = "6",
    pages = "062005",
    year = "2024"
}

@article{PierreAuger:2025lsh,
    author = "{D. Schmidt for the \textsc{Pierre Auger} Collaboration}",
    title = "{AugerPrime: Status and first results}",
    eprint = "2508.08056",
    archivePrefix = "arXiv",
    primaryClass = "astro-ph.IM",
    reportNumber = "PoS-ICRC2025-385",
    doi = "10.22323/1.501.0385",
    journal = "PoS",
    volume = "ICRC2025",
    pages = "385",
    year = "2025"
}

@misc{Denise_Boncioli,
author = "Boncioli, Denise",
    title =  "private communication",
    year = "2026",
}

@article{Alhebsi:2026bdk,
    author = "Alhebsi, Abdulrahman and van Vliet, Arjen and Ehlert, Domenik and Thoudam, Satyendra",
    title = "{Implications of a Cosmogenic Origin of KM3-230213A for Ultra-high-energy Protons}",
    eprint = "2603.13476",
    archivePrefix = "arXiv",
    primaryClass = "astro-ph.HE",
    doi = "10.3847/1538-4357/ae6452",
    journal = "Astrophys. J.",
    volume = "1003",
    number = "2",
    pages = "235",
    year = "2026"
}

@article{Muzio:2019leu,
    author = "Muzio, Marco Stein and Unger, Michael and Farrar, Glennys R.",
    title = "{Progress towards characterizing ultrahigh energy cosmic ray sources}",
    eprint = "1906.06233",
    archivePrefix = "arXiv",
    primaryClass = "astro-ph.HE",
    doi = "10.1103/PhysRevD.100.103008",
    journal = "Phys. Rev. D",
    volume = "100",
    number = "10",
    pages = "103008",
    year = "2019"
}

@article{Dominguez:2010bv,
    author = "Dominguez, Alberto and others",
    title = "{Extragalactic Background Light Inferred from AEGIS Galaxy SED-type Fractions}",
    eprint = "1007.1459",
    archivePrefix = "arXiv",
    primaryClass = "astro-ph.CO",
    doi = "10.1111/j.1365-2966.2010.17631.x",
    journal = "Mon. Not. Roy. Astron. Soc.",
    volume = "410",
    pages = "2556",
    year = "2011"
}

@article{Stecker:2005qs,
    author = "Stecker, Floyd W. and Malkan, Matthew A. and Scully, Sean T.",
    title = "{Intergalactic photon spectra from the far ir to the uv lyman limit for 0 {\ensuremath{<}} Z {\ensuremath{<}} 6 and the optical depth of the universe to high energy gamma-rays}",
    eprint = "astro-ph/0510449",
    archivePrefix = "arXiv",
    doi = "10.1086/506188",
    journal = "Astrophys. J.",
    volume = "648",
    pages = "774--783",
    year = "2006"
}

@article{Berezinsky:2016feh,
    author = "Berezinsky, V. and Kalashev, O.",
    title = "{High energy electromagnetic cascades in extragalactic space: physics and features}",
    eprint = "1603.03989",
    archivePrefix = "arXiv",
    primaryClass = "astro-ph.HE",
    doi = "10.1103/PhysRevD.94.023007",
    journal = "Phys. Rev. D",
    volume = "94",
    number = "2",
    pages = "023007",
    year = "2016"
}

@article{Fermi-LAT:2014ryh,
    author = "Ackermann, M. and others",
    collaboration = "Fermi-LAT",
    title = "{The spectrum of isotropic diffuse gamma-ray emission between 100 MeV and 820 GeV}",
    eprint = "1410.3696",
    archivePrefix = "arXiv",
    primaryClass = "astro-ph.HE",
    doi = "10.1088/0004-637X/799/1/86",
    journal = "Astrophys. J.",
    volume = "799",
    pages = "86",
    year = "2015"
}

\end{document}